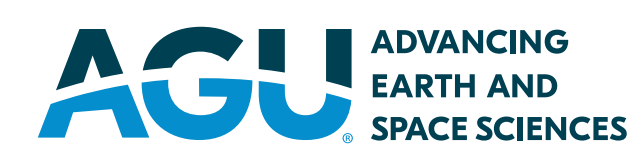

**Key Points:**

- The resonant interaction of whistler-mode chorus and relativistic electrons in the magnetosphere can be modeled as a free-electron laser
- We derive a closed set of nonlinear equations for the time evolution of the system in terms of just three collective variables
- The spatially inhomogeneous, multi-mode wave amplitude is approximated by the Ginzburg-Landau equation, predicting solitary whistler waves

**Correspondence to:**
B. Bonham,
bbonham@princeton.edu

**Author Contributions:**
**Conceptualization:** Brandon Bonham, Amitava Bhattacharjee
**Formal analysis:** Brandon Bonham
**Funding acquisition:** Brandon Bonham, Amitava Bhattacharjee
**Investigation:** Brandon Bonham, Amitava Bhattacharjee
**Project administration:** Amitava Bhattacharjee
**Software:** Brandon Bonham
**Supervision:** Amitava Bhattacharjee
**Validation:** Brandon Bonham, Amitava Bhattacharjee
**Visualization:** Brandon Bonham

# Whistler Chorus Amplification in the Magnetosphere: The Nonlinear Free-Electron Laser Model and the Ginzburg-Landau Equation

**Brandon Bonham[1] and Amitava Bhattacharjee[2]**

[1]Department of Physics, Princeton University, Princeton, NJ, USA, [2]Department of Astrophysical Sciences, Princeton University, Princeton, NJ, USA

**Abstract** We present a novel nonlinear model for whistler-mode chorus amplification based on the free-electron laser (FEL) mechanism. First, we derive the nonlinear collective variable equations for the whistler-electron interaction. Consistent with in situ satellite observations, these equations predict that a small seed wave can undergo exponential growth, reaching a peak of a few hundred picoteslas after a few milliseconds, followed by millisecond timescale amplitude modulations. Next, we show that when one accounts for multiple wave frequencies and wave spatial variations, the amplitude and phase of the whistler wave can be described by the Ginzburg-Landau equation (GLE), providing a framework for the investigation of solitary wave behavior of chorus modes. These findings enhance our understanding of wave-particle interactions and space weather in the Van Allen radiation belts, deepen the connection between whistler-electron dynamics and FELs, and reveal a novel connection between whistler-mode chorus and the GLE.

**Plain Language Summary** Chorus waves are electromagnetic waves named for their resemblance to birds chirping at dawn when their radio frequencies are played as audio. In the Earth's Van Allen radiation belts, interactions between chorus waves and electrons can strongly amplify these waves, influencing space weather and posing dangers to satellites and probes traversing the region. Recent efforts to understand chorus wave amplification have drawn upon parallels to free-electron lasers, laboratory devices that generate intense laser beams with tunable frequencies. This approach is known as the free-electron laser model. In this study, we reveal new insights about the long-term behavior predicted by this model and new agreements between the model and observational data. This work also proposes that when multiple chorus waves of varying frequencies coexist, their behavior is governed by one of the most-studied nonlinear equations in physics, the Ginzburg-Landau equation, resulting in the prediction of solitary waves. This phenomenon, first noticed in the context of water waves propagating in a canal, is characterized by isolated waves traveling for surprising distances without changing their shape or speed. Overall, these findings offer new perspectives on chorus waves and could inform strategies to protect space-based technologies from radiation belt hazards.

## 1. Introduction

Planetary radiation belts contain two types of whistler emissions - chorus and hiss - which are primarily generated near the magnetic equator (see the textbook by Gurnett and Bhattacharjee (2017) for an introduction). Hiss consists of a broad, noisy spectrum and plays a major role in radiation belt particle loss (Agapitov et al., 2020; Ripoll et al., 2017). Chorus is characterized by discrete frequency modes that often sweep rapidly upwards and/or downwards in frequency - a phenomenon known as chirping. Resonant wave-particle interactions can cause amplification of the waves from a few picoteslas to hundreds of picoteslas and acceleration of the electrons from hundreds of KeV to tens of MeV or higher (Mourenas et al., 2023; Summers & Stone, 2022). This high-energy radiation poses risks to satellites and other spacecraft transiting the region, making it essential to understand the amplification mechanisms that govern these waves.

The whistler wave amplification process has been the subject of several theoretical models (see Hanzelka and Santolík (2024) for a recent review). Recently, it was realized that the dynamics of electrons in the rotating wave field of a whistler can be analogized to the dynamics of electrons in the radiation field of a free-electron laser (FEL). For a brief and accessible introduction to FELs, see Margaritondo and Rebernik Ribic (2011). More comprehensive treatments can be found in the monographs by Marshall (1985), Brau (1990), Freund and Antonsen (2024). The basic idea behind FELs is the following. A periodic arrangement of permanent magnets

**Writing – original draft:** Brandon Bonham
**Writing – review & editing:** Brandon Bonham, Amitava Bhattacharjee

with opposing polarities is used to generate a static sinusoidal magnetic field known as a wiggler or undulator. A beam of relativistic electrons is injected into the wiggler field, causing them to oscillate in the direction transverse to the beam line and emit electromagnetic waves. The Lorentz force due to the B-field of these waves causes the electrons to form bunches which emit increasingly coherent, therefore intense, radiation.

The analogy between whistler wave amplification in the magnetosphere and the FEL amplification mechanism was proposed by Soto-Chavez et al. (2012), who derived dynamical equations for the magnetospheric system which have the form of FEL equations. In this formulation, the role of the undulator magnetic field in a FEL is played by whistler waves in the plasma (produced by another mechanism such as anisotropic electron distribution functions), but amplified by the FEL mechanism. The role of the FEL electron beam is played by relativistic electrons in the radiation belt which are assumed to be mono-energetic. Using a collective variable approach, Soto-Chavez et al. (2012) derived an analytic expression for the linear growth rate of the wave amplitude and provided estimates for the saturation amplitude and timescale. More recently, the model of Zonca et al. (2022), strengthened the FEL analogy by drawing a direct connection between chorus chirping and FEL superradiance. Interestingly, the FEL mechanism has also recently been used to model phenomena in other astrophysical contexts, such as pulsars, magnetars, and fast-radio bursts (Fung & Kuijpers, 2004; Lyutikov, 2021).

In this study, we investigate the nonlinear regime of the FEL model of chorus amplification. We employ the model to better understand the nonlinear structure of the whistler wave packets that are often observed to show a high degree of coherence, and to investigate whether these wave packets are solitary wave-like structures (or even more robust structures such as solitons which survive the effect of collisions). These questions are of theoretical as well as observational interest, discussed later in the paper.

In the original work of Soto-Chavez et al. (2012), the authors derived $2N + 2$ dynamical equations for the whistler-electron interaction, and reduced these to a set of just three linear equations written in terms of collective variables. Here, we derive the nonlinear collective variable equations, enabling analytical predictions of the saturation amplitude and post-saturation behavior of the wave. We then show that the exponential growth phase and mean saturation behavior can be modeled by the Stuart-Landau equation (SLE) - a simple, yet universal equation for oscillators with a weak nonlinearity (García-Morales & Krischer, 2012). In addition, by generalizing to the case of wave spatial dependence and multiple frequencies, we show that the multi-mode wave amplitude is governed approximately by a Ginzburg-Landau equation (GLE), which is one of the most celebrated nonlinear equations in physics.

The GLE has a broad range of applications both within and outside fluid mechanics and plasma physics, having been applied to phenomena such as superfluidity and superconductivity, liquid crystals, and chaotic spirals in slime molds (Aranson & Kramer, 2002; García-Morales & Krischer, 2012). The GLE is known to admit special solitary wave solutions (Nozaki & Bekki, 1984) - localized nonlinear wave packets that maintain a definite shape despite being composed of a spectrum of wavelengths with different phase velocities (Craik, 2004). They are closely related to solitons, but they lack the elastic scattering property that allows solitons to maintain their shape and speed after colliding with each other. In summary, this work investigates the nonlinear behavior of whistler-mode chorus using the FEL model, strengthens the connection between radiation belt physics and the well-established field of FELs, and provides a framework for the investigation of solitary chorus modes through a novel connection to the GLE.

## 2. Theory

### 2.1. Derivation of Nonlinear Collective Variable Equations

We begin with Equations 1–3, as derived by Soto-Chavez et al. (2012), which describe a monochromatic whistler wave interacting with $N$ relativistic electrons near the geomagnetic equator. The whistler is assumed to have the form $\mathbf{B}_w = B_w(t)(\hat{x}\cos\varphi + \hat{y}\sin\varphi)$, where $\varphi = \omega(k)t - kz + \phi(t)$, $\omega(k)$ (hereafter denoted $\omega$) is given by the cold plasma dispersion relation, $k$ is the wave number, $z$ is the axis along the background geomagnetic field, and $\phi(t)$ (hereafter denoted $\phi$) is a time-dependent phase. The electrons, assumed to have a constant velocity perpendicular to the background field ($v_\perp = v_{\perp 0}$), are coupled to the whistler field by a complex current in Maxwell's equations, and their change in momentum along the field is given by the Lorentz force equation. Using Maxwell's equations with a slowly varying amplitude and phase approximation the whistler amplitude and phase can be combined into a single complex equation. Ultimately, at the magnetic equator, the closed set of equations for the electron-wave phase difference, the electron proper velocity parallel to the background field, and the complex wave field respectively are.

$$\frac{d\psi_j}{dt} = \frac{\Omega_{e0}}{\gamma_j} - \omega + k\frac{\eta_z^j}{\gamma_j} \quad (j = 1 \ldots N) \tag{1}$$

$$\frac{d\eta_z^j}{dt} = \frac{v_{\perp 0}}{2}\left(ibe^{-i\psi_j} + \text{c.c.}\right) \quad (j = 1 \ldots N) \tag{2}$$

$$\frac{db}{dt} = i\frac{sv_{\perp 0}\Omega_{e0}\omega_{pr}^2}{2kc^2}\frac{1}{N}\sum_{j=1}^{N} e^{i\psi_j}. \tag{3}$$

These are in the form of the well-known equations for a FEL. Equation 1 governs the time evolution of $\psi = \theta - \omega t + kz = \theta - \varphi + \phi$, where $\theta$ is the electron gyro-phase with respect to the $x$-axis in the plane perpendicular to the background field. Hence, $\psi_j$ is related to the $j$th electron's phase with respect to the whistler wave field. The parameter $\Omega_{e0}$ is the electron cyclotron frequency due to the background field at the equator, $\eta_z^j = \gamma_j v_z^j$ is the parallel proper velocity of the $j$th electron, and $\gamma_j = \left(1 + \vec{\eta}_j^{\,2}/c^2\right)^{1/2}$ is the Lorentz factor for the $j$th electron. Equation 2 describes the time evolution of the parallel proper velocity of the $j$th electron. The variable $b \equiv (e/m_e)B_w e^{i\phi}$ is the scaled complex magnetic field of the whistler, where $e$ is the magnitude of the charge of the electron, and $m_e$ is the mass of the electron. The original equation includes an additional term that accounts for inhomogeneities in the background dipole field as a function of the distance above the magnetic equator. Since chorus amplification is known to take place within a few degrees of this plane (Santolík et al., 2003), in this work we focus on the magnetic equator where this term vanishes. Last, Equation 3 describes the time evolution of the complex whistler field, where $s = \frac{\omega}{\Omega_{e0} - \omega}$, $\omega_{pr}$ is the electron plasma frequency of the $N$ resonant electrons, and $c$ is the speed of light.

Following Bonifacio et al., 1986 one can define collective variables which can be used to reduce the above $2N + 2$ equations to just three (Bonifacio et al., 1986). Here, the collective variables are defined as,

$$A \equiv be^{-i\psi_0} \tag{4}$$

$$X \equiv \frac{1}{N}\sum_{j=1}^{N} e^{i\left(\psi_j - \psi_0\right)} \equiv \langle e^{i\Delta\psi}\rangle \tag{5}$$

$$Y \equiv \frac{1}{N}\sum_{j=1}^{N} e^{i\left(\psi_j - \psi_0\right)}\left(\eta_z^j - \eta_{z0}\right) \equiv \langle \Delta\eta_z e^{i\Delta\psi}\rangle \tag{6}$$

where $\langle(\ldots)\rangle \equiv \frac{1}{N}\sum_{j=1}^{N}(\ldots)$ is the average over all electrons, $\eta_{z0} = \gamma_0 v_{z0}$ is the parallel proper velocity of each electron in the initially mono-energetic electron beam, $\gamma_0$ is the initial Lorentz factor, $\Delta\eta_z^j \equiv \eta_z^j - \eta_{z0}$, and $\Delta\psi_j \equiv \psi_j - \psi_0$. Note, the parameter $\psi_0$ is not a constant, but increases linearly with time according to its definition via the detuning constant $\delta \equiv d\psi_0/dt = \Omega_{e0}/\gamma_0 - \omega + k\eta_{z0}/\gamma_0$. The detuning parameter can be written more instructively as $\delta = k\left(v_{z0} - v_r\right)$, where $v_r \equiv (\omega - \Omega_{e0}/\gamma)/k$ is the electron resonance velocity.

The collective variables, $A$, $X$, and $Y$, can be thought of as amplitude, phase, and momentum variables respectively. In particular, $|A| = (e/m_e)B_w$ is proportional to the amplitude of the whistler. The variable $X$, known in the FEL literature as the bunching variable, measures the degree of randomness of the electron phases. As can be seen from the definition, if the electron phases are random, then $X$ tends to zero, and if the electrons all have the same phase, then $|X| = 1$. The variable $Y$, being a mixed variable has a less direct interpretation, but plays the role of a phase weighted momentum.

To obtain the collective variable form of Equations 1–3 we take time derivatives of the collective variable definitions, (4)–(6), and replace terms containing $\dot{\psi}$,$\dot{\eta}_z$, and $\dot{b}$ with their definitions in the original wave-particle equations, (1)–(3). We also expand $\gamma$ and $1/\gamma$ about the initial proper velocity, that is $\gamma = \gamma_0 + \Gamma_1 \Delta\eta_z + (1/2)\Gamma_2 \Delta\eta_z^2 + O\left(\Delta\eta_z^3\right)$ and $1/\gamma = \gamma_0^{-1} + \Gamma_{-1}\Delta\eta_z + (1/2)\Gamma_{-2}\Delta\eta_z^2 + O\left(\Delta\eta_z^3\right)$, where $\Gamma_n \equiv \frac{d^n\gamma}{d\eta_z^n}\Big|_{\eta_{z0}}$. Since $v_{\perp 0}$ is a constant, one may write $\gamma = \gamma_{\perp 0}\left(1 + \eta_z^2/c^2\right)^{1/2}$, where $\gamma_{\perp 0} \equiv \left(1 - v_{\perp 0}^2/c^2\right)^{-1/2}$, to obtain each $\Gamma_n$. To order $\langle \Delta\eta_z^2 e^{i\Delta\psi}\rangle$ the collective variable equations are,

$$\dot{A} = igX - i\delta A \tag{7}$$

$$\dot{X} \simeq -ih_c Y \tag{8}$$

$$\dot{Y} \simeq iuA - ih_c\langle \Delta\eta_z^2 e^{i\Delta\psi}\rangle \tag{9}$$

where $h_c \equiv -\left(\Gamma_{-1}\Omega_{e0} + k/\gamma_0 + \Gamma_{-1}k\eta_{z0}\right) = -\frac{k}{\gamma_0}\left(1 - \frac{\gamma_{\perp 0}^2\eta_{z0}^2}{\gamma_0^2 c^2} - \frac{\gamma_{\perp 0}^2\Omega_{e0}\eta_{z0}}{\gamma_0^2 c^2 k}\right)$, $u \equiv v_{\perp 0}/2$, $g \equiv \left(s\Omega_{e0}\omega_{pr}^2 u/kc^2\right)$, and the overdots indicate $\frac{\mathrm{d}}{\mathrm{d}t}$. Note the term $i\zeta_0\langle \Delta\eta_z^2 e^{i\Delta\psi}\rangle$, where $\zeta_0 \equiv \frac{1}{2}\Gamma_{-2}\Omega_{e0} + \Gamma_{-1}k + \frac{1}{2}\Gamma_{-2}k\eta_{z0}$, has been omitted in Equation 8 since $h_c/\zeta_0 \sim 10^{-10}$ for typical magnetospheric parameters such as those considered here. Last, a term $-iuA^*\langle e^{2i\Delta\psi}\rangle$ has been omitted in Equation 9 since it has been shown to be negligible compared to the remaining nonlinear term (Bonifacio et al., 1986).

To obtain a closed set of collective variable equations one must eventually apply a closure condition, else continually define higher orders of $\langle \Delta\eta_z^n e^{i\Delta\psi}\rangle$. If we assume moments higher than $n = 1$ vanish, then we obtain the linear collective variable equations of the FEL model (Soto-Chavez et al., 2012). To obtain the corresponding nonlinear equations, following Bonifacio et al. (1986), we obtain closure by assuming all moments higher than $n = 2$ vanish, and employing the factorization assumption,

$$\begin{aligned} &\left\langle \left(\Delta\eta_z - \langle\Delta\eta_z\rangle\right)^2 e^{i\Delta\psi}\right\rangle \simeq \left\langle \left(\Delta\eta_z - \langle\Delta\eta_z\rangle\right)^2\right\rangle \langle e^{i\Delta\psi}\rangle \\ \Rightarrow\ &\langle \Delta\eta_z^2 e^{i\Delta\psi}\rangle \simeq X\left(\langle\Delta\eta_z^2\rangle - 2\langle\Delta\eta_z\rangle^2\right) + 2Y\langle\Delta\eta_z\rangle \end{aligned} \tag{10}$$

The system admits the following conservation relations which can be used to determine $\langle\Delta\eta_z\rangle$ and $\langle\Delta\eta_z^2\rangle$,

$$P_0 = \langle\Delta\eta_z\rangle + \frac{u}{g}|A|^2 \tag{11}$$

$$H_0 = \frac{\langle\Delta\eta_z^2\rangle}{2} - \frac{u}{h_c}\left(A^*X + X^*A\right) + \frac{u\delta}{gh_c}|A|^2 \tag{12}$$

where $P_0$ and $H_0$ are constants. The first equation is an exact relation of Equations 1–3, and reflects the conservation of momentum. The second is an exact relation of the nonlinear collective variable Equations 7–9, and is associated with the Hamiltonian for the system. One obtains,

$$\begin{aligned} \langle \Delta\eta_z^2 e^{i\Delta\psi}\rangle \simeq &\left(2H_0 - 2\frac{u^2}{g^2}|A_0|^4\right)X + 2\frac{u}{g}|A_0|^2 Y \\ &+\left(4\frac{u^2}{g^2}|A_0|^2 - 2\frac{u\delta}{gh_c}\right)|A|^2X - 2\frac{u}{g}|A|^2Y + 2\frac{u}{h_c}\left(A^*X + X^*A\right)X - 2\frac{u^2}{g^2}|A|^4X \end{aligned} \tag{13}$$

The two linear terms constitute a small correction to the exponential growth regime and can be dropped. Similarly, one may omit the first term in the coefficient of $|A|^2X$. Finally, substituting the expression into Equations 7–9 yields the nonlinear collective variable equations for the FEL model of chorus amplification,

$$\dot{A} = igX - i\delta A \tag{14}$$

$$\dot{X} \simeq -ih_c Y \tag{15}$$

$$\dot{Y} \simeq iuA + 2i\delta\frac{u}{g}|A|^2X + 2ih_c\frac{u}{g}|A|^2Y - 2iu(A^*X + X^*A)X + 2ih_c\frac{u^2}{g^2}|A|^4X. \tag{16}$$

Given the constants, $g, \delta, u,$ and $h_c$, this simple set of equations allows one to calculate the nonlinear time evolution of the whistler amplitude and phase, and the mean electron momentum and phase.

### 2.2. Ginzburg-Landau Equation

Thus far, we have considered the dynamical behavior of the system in the presence of a single-frequency mode, ignoring its spatial structure. In the presence of spatially varying modes with multiple frequencies it is difficult to obtain a closed set of collective variable equations such as the ones above. However, it was first shown by Cai and Bhattacharjee (1991) by an analogy between the electron beam and an optical fiber that the wave amplitude in the collective variable equations for a FEL can be modeled by a GLE. The analogous equations for chorus amplification (14)–(16) are of the same form, therefore can also be modeled by a GLE. The derivation of the GLE, sketched here, is given most clearly in (Ng & Bhattacharjee, 1998) for the case of FELs. First, substituting the linear order solutions to the collective variable equations into the third order expression for Equation 16 yields,

$$\dot{A} \simeq i\lambda_0 A + i\beta|A|^2A \tag{17}$$

where $\lambda_0$ is the dominant root of the characteristic cubic equation $\left(\lambda_0^3 + \delta\lambda_0^2 + ugh_c = 0\right)$ of the linearized collective variable Equations 14–16, and the constant $\beta$ is given by,

$$\beta = -2uh_c\lambda_0\left(-\frac{1}{g\lambda_0} + \frac{\delta}{g\lambda_0^2} + \frac{uh_c}{\lambda_0^4} + \frac{uh_c}{|\lambda_0|^4}\right). \tag{18}$$

Equation 17 is the SLE mentioned above. It provides a simplified model for the amplitude behavior, describing initial exponential growth followed by saturation at a value of $\langle|A_{\text{sat}}|\rangle = \sqrt{-\lambda_i/\beta_i}$, where $\lambda_r + i\lambda_i \equiv \lambda_0$ and $\beta_r + i\beta_i \equiv \beta$.

If we consider the behavior of a whistler wave packet with a continuous spectrum of frequencies centered near a reference frequency $\omega_0$, and allow for spatial variation in the wave amplitude, we obtain the GLE for the behavior of chorus waves according to the FEL model,

$$v_g\frac{\partial A}{\partial z} = i\lambda_0(\omega_0)A - \mu\frac{\partial A}{\partial t} - i\frac{\alpha}{2}\frac{\partial^2 A}{\partial t^2} + i\beta|A|^2A. \tag{19}$$

Where $A = A(z,t)$, $v_g = \partial z/\partial t = (\partial\omega/\partial k)|_{\omega_0}$, $\lambda_0(\omega_0)$ is the linear growth rate evaluated at $\omega_0$, $\mu \equiv 1 + \lambda_0'(\omega_0)$, and $\alpha \equiv \lambda_0''(\omega_0)$, where the primes indicate derivatives with respect to $\omega$. For convenience, we define $\mu_r + i\mu_i \equiv \mu$, $\alpha_r + i\alpha_i \equiv \alpha$, denote division by the group velocity with an overbar, for example $\bar{\lambda}_r \equiv \lambda_r/v_g$, and define $c_1 = -\alpha_r/\alpha_i$ and $c_2 = -\beta_r/\beta_i$. Finally, by applying the transformations $A(z,t) = \Phi_0\Phi(\zeta,\tau)\exp i[K_0\zeta + \Omega_0\tau]$, $\zeta = z/z_0$, $\tau = t/t_0 + z/v_0t_0$, where,

$$\begin{aligned} z_0 &= 1/\left(-\bar{\lambda}_i + \bar{\mu}_i^2/2\bar{\alpha}_i\right), & t_0^2 &= \bar{\alpha}_i z_0/2, \\ v_0 &= 1/\left(-\bar{\mu}_r + \bar{\alpha}_r\bar{\mu}_i/\bar{\alpha}_i\right), & \Phi_0^2 &= 1/z_0\bar{\beta}_i, \\ K_0 &= z_0\left(\bar{\lambda}_r - \bar{\alpha}_r\bar{\mu}_i^2/2\bar{\alpha}_i^2\right), & \Omega_0 &= \bar{\mu}_i t_0/\bar{\alpha}_i, \end{aligned} \tag{20}$$

one obtains the GLE in standard form,

$$\frac{\partial\Phi}{\partial\zeta} = \Phi + (1 + ic_1)\frac{\partial^2\Phi}{\partial\tau^2} - (1 + ic_2)|\Phi|^2\Phi. \tag{21}$$

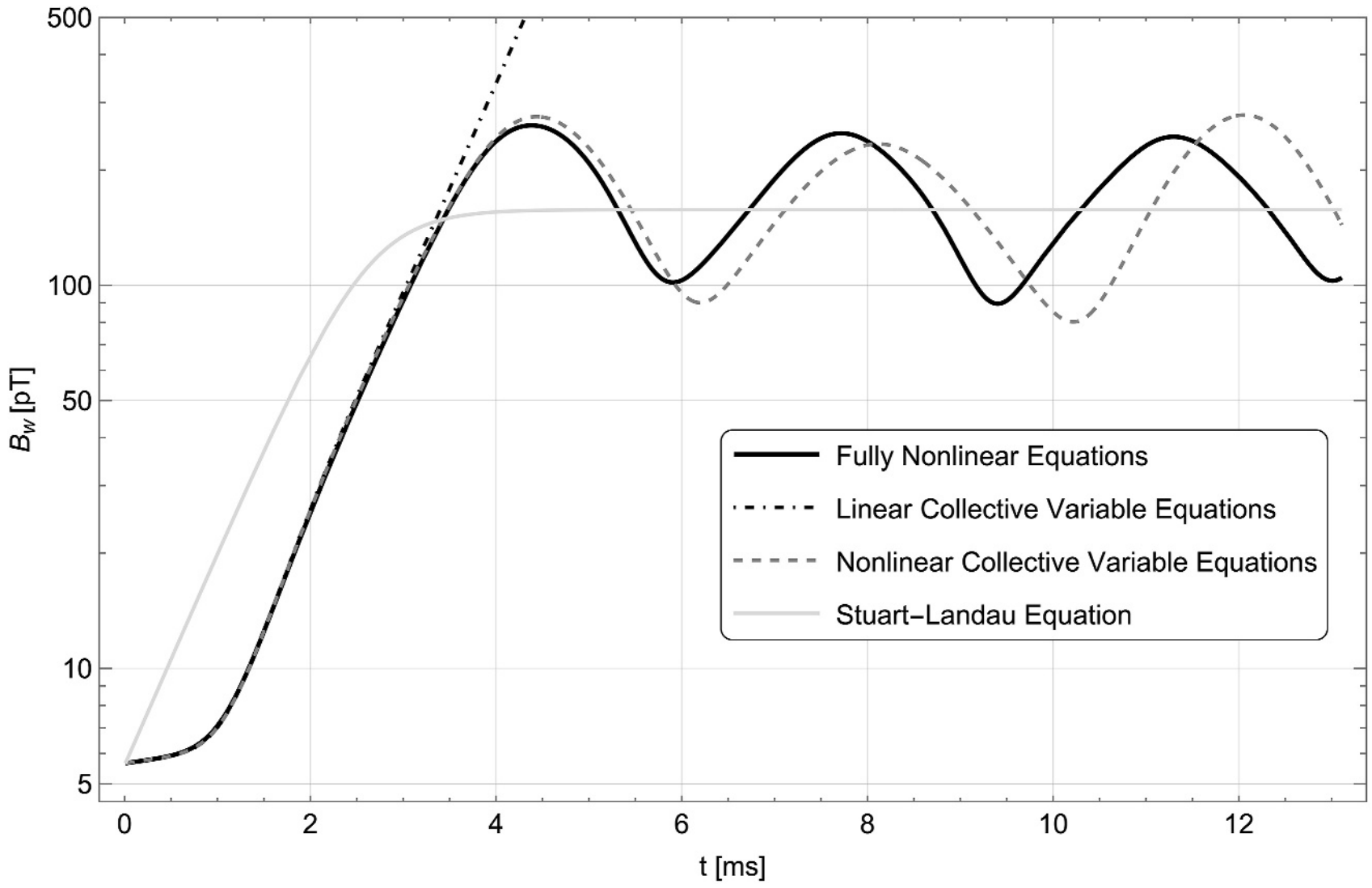

**Figure 1.** A comparison between the fully nonlinear original $2N + 2$ dynamical equations (Equations 1–3), the linear collective variable equations (Linearized Equations 14–16) appearing in the work of Soto-Chavez et al. (2012), the nonlinear collective variable equations (Equations 14–16), and the Stuart-Landau Equation (Equation 17) for typical magnetospheric conditions. The nonlinear collective variable equations agree with the linear collective variable equations in the exponential growth regime, and agree with the fully-nonlinear equations in the post-saturation regime.

## 3. Discussion

In the single-mode case, the general behavior of the wave amplitude is exponential growth followed by saturation and post-saturation oscillations. This is shown in Figure 1 for typical magnetospheric conditions, such as those used in the original FEL model analysis by Soto-Chavez et al. (2012). For completeness, we restate the conditions here: a whistler with a frequency of $\omega_0 = 0.3\Omega_{e0}$ (near the resonance frequency occurring at $\delta = 0$), wavelength of 8,131 km, and initial amplitude of 5.7 pT interacting with a monoenergetic beam of electrons with initial parallel velocities of $v_{z0} = -0.157c$, constant perpendicular velocities of $v_{\perp 0} = 0.68c$, initial bunching parameter of $X_0 \approx 0.001$, and plasma frequency of $\omega_{pr} = 10^3$rads/s located at the magnetic equator on the $L = 4$ shell where the background field is 0.5 μT.

Figure 1 shows exponential growth at a rate of $|\lambda_i| \sim 10^3$ s$^{-1}$ for a few milliseconds with a saturation amplitude around 250 pT followed by amplitude modulations of width ~2 ms and period ~3.5 ms. Away from the narrow resonance at $\delta = 0$, the linear growth rates are of the order of hundreds of s$^{-1}$. These predictions are in good agreement with in situ satellite measurements, which indicate exponential growth at a rate of up to a few hundred s$^{-1}$ followed by a peak amplitude at typical values of a few hundred pT (Santolik, 2008). Furthermore, these measurements indicate that amplitude peaks can be followed by periods of decay and further growth, resulting in amplitude oscillations with a duration of a few milliseconds to a few tens of milliseconds. Such amplitude modulations are common, having been observed in many other cases (Dubinin et al., 2007; Li et al., 2011; Mozer et al., 2021; Santolík et al., 2014). Last, one of the dominant motivations for the development of this laser-like model is the well known fact that chorus modes are highly phase coherent (Agapitov et al., 2017), which is attributed to their resonant interaction with phase bunched (coherent) electrons. This is reflected in our model by the phase bunching parameter $X$ growing from it's initially near-random state to a peak value around 0.8 in lockstep with the amplitude growth.

The multi-mode spatially-varying chorus behavior is governed approximately by the GLE. The bandwidth of the resulting wave packets can be estimated using the parabolic approximation for the growth rate employed in the derivation of the GLE, resulting in the expression, $\delta\omega \approx 2\sqrt{-2\lambda_i/\alpha_i}$, or $\delta\omega \approx 0.05\ \Omega_{e0}$ for the parameters considered here. This is consistent with CLUSTER spacecraft observations, which indicate the majority of chorus wave packets have a bandwidth in the range of 0.03 $\Omega_{e0}$ to 0.2 $\Omega_{e0}$, with lower values occurring more often near the magnetic equator (Santolik et al., 2008). Furthermore, it is well known that the GLE admits special solitary wave solutions. This predicts that, despite being composed of a spectrum of frequencies - each with distinct growth rates and phase velocities - the amplitude envelope of a wave packet can propagate with a definite shape and velocity, as first described by Russell (1844) for the case of shallow water waves. Such solutions have been derived in general (Cariello & Tabor, 1989; Nozaki & Bekki, 1984), and discussed more particularly for the case of free-electron lasers (Cai & Bhattacharjee, 1991). In the latter case, in which the GLE is derived from collective variable equations of the same form as appear here, the pulse widths have been compared favorably to data from several FEL experiments. The solitary wave solutions to Equation 21 are pulses of the form,

$$\Phi = \frac{Qe^{-i\Omega\zeta}}{(2\cosh K\tau)^{1+i\sigma}}, \tag{22}$$

where,

$$|Q| = 2\left(1 + \frac{c_1\sigma - 1}{\sigma^2 + 2c_1\sigma - 1}\right)^{1/2}, \Omega = c_1 - \frac{2\sigma(1+c_1)}{\sigma^2 + 2c_1\sigma - 1}, K = \left(\frac{1}{\sigma^2 + 2c_1\sigma - 1}\right)^{1/2}, \tag{23}$$

and $\sigma$ satisfies the quadratic equation,

$$\sigma^2 - 3\frac{c_1c_2 + 1}{c_1 - c_2}\sigma - 2 = 0. \tag{24}$$

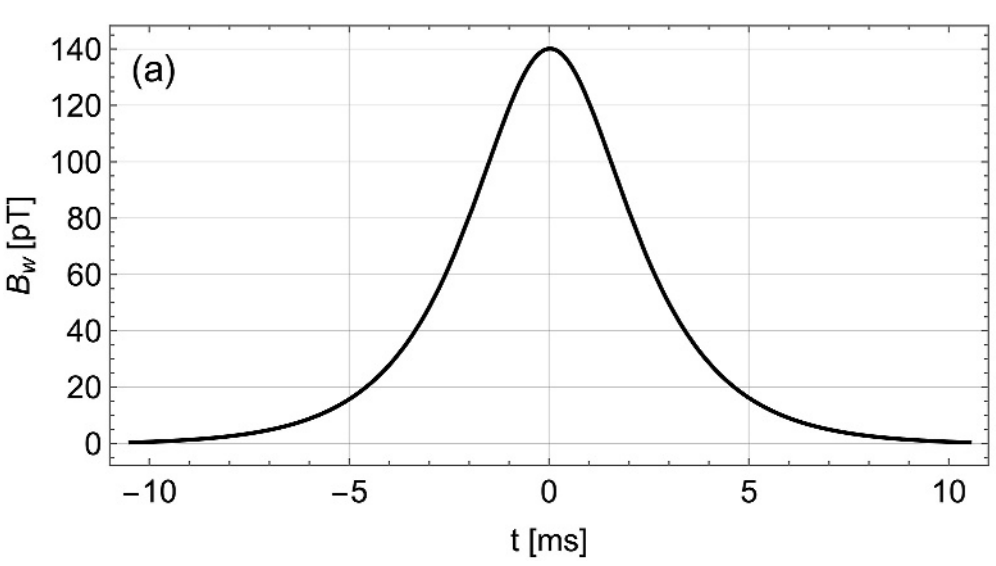

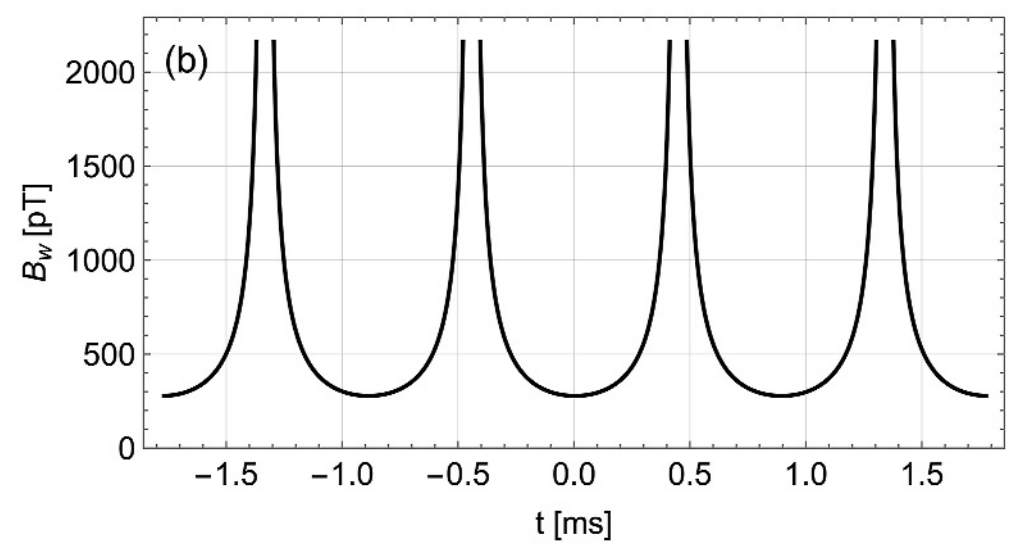

**Figure 2.** Solitary wave solutions of the Ginzburg-Landau equation (Equation 19). (a) The solitary pulse corresponding to Equation 26. (b) The periodic train of solitary waves with finite-time singularities corresponding to Equation 27.

By transforming back to the original field and coordinates, we obtain the solitary wave solutions to Equation 19 in physical space,

$$A = \frac{\Phi_0 Q}{\left[2\cosh K\left(\frac{t}{t_0} + \frac{z}{v_0t_0}\right)\right]^{1+i\sigma}} \exp\left\{i\left[z\left(\frac{K_0}{z_0} + \frac{\Omega_0}{v_0t_0} - \frac{\Omega}{z_0}\right) + t\left(\frac{\Omega_0}{t_0}\right)\right]\right\}. \tag{25}$$

For the magnetospheric parameters considered here, the roots of the equation for $\sigma$ are real, but result in one purely real $K$ and one purely imaginary $K$. In the former case, where $K = K_r \in \mathbb{R}$, at the origin one obtains the single pulse,

$$|A(t)| = \frac{|\Phi_0 Q|}{2\cosh(K_r t/t_0)}. \tag{26}$$

In the latter case, where $K = K_i \in \mathbb{I}$, at the origin one obtains multiple pulses with periodic singularities,

$$|A(t)| = \frac{|\Phi_0 Q|}{|2\cos(|K_i| t/t_0)|}, \tag{27}$$

where we have used the approximation $|\sigma| \ll 1$ (Figure 2). The single pulse is gaussian-like, with a width given by $\Delta t = 2t_0/K_r$, which is approximately $\Delta t \simeq \sqrt{-2\alpha_i\sigma^2/\lambda_i}$ in the limit $\sigma \gg 1$. Substituting the parameters used here yields a peak amplitude of 140 pT and a width of $\Delta t \approx 3.5$ ms. In contrast, the singular periodic solution consists of a train of narrow spikes, with widths given by $\Delta t = 2t_0/|K_i| \approx 0.5$ ms, which is approximately $\Delta t \simeq \sqrt{-2\alpha_i/\lambda_i}$ in the limit $\sigma \ll 1$. The separation between neighboring singularities is given by $\Delta T = \pi t_0/|K| \approx 0.9$ ms, which is approximately $\Delta T \simeq \pi\sqrt{-\alpha_i/2\lambda_i}$ for $\sigma \ll 1$. Although the validity of the periodic solution (with its finite-time singularities) is limited and the solutions will be regularized by effects outside the realm of validity of our model, it nevertheless predicts a qualitatively similar sequence of millisecond timescale amplitude spikes and a tendency for spike formation. In fact, both solutions exhibit modulation timescales comparable to those of the single-mode case, in agreement with the observational evidence discussed above. Furthermore, in the context of magnetospheric lion roars, Dubinin et al. (2007) reported that bursts of whistler emissions - also excited by a local population of resonant electrons - exhibit amplitude variations consistent with solitary wave behavior. The associated wave turbulence was found to consist of many nearly monochromatic, circularly polarized wave packets, with field strength variations resembling solitary structures. Given that chorus arises from a similar mechanism, it is plausible that chorus modes also exhibit such solitary wave behavior.

Finally, we note some limitations of the present model and possible directions for future research. First, the single-mode treatment - which results in the reduced nonlinear equations (Equations 14–16) - assumes a whistler wave with a fixed frequency and a time-dependent phase. Despite the tendency of chorus modes to chirp in frequency even in the absence of background inhomogeneities (Omura & Nunn, 2011), here we do not investigate the physics of the frequency variations contained within the time-dependent phase. In general, a multi-mode approach, similar to that employed in our derivation of the GLE, may also be capable of modeling chirping behavior. However, since the SLE averages out the post-saturation oscillations indicative of resonant electron trapping - one of the mechanisms associated with frequency chirping - the degree to which the GLE, as presented here, can model chorus chirping is uncertain. The investigation of chirping in the context of the FEL model is a promising direction for future research. Since the FEL analogy works in both directions, such an investigation could also provide insight into frequency chirping in free-electron lasers. Next, while Equations 14–16 offer a straightforward route to the analysis of both wave and particle behavior, this work focuses primarily on the wave behavior, leaving the investigation of the model's implications for particle dynamics to future work. In addition, we note that our derivation of the GLE, being based on an extension of single-mode spatially-independent equations could perhaps be derived more directly as a limiting case of a more general treatment. Last, we note the interesting possibility for the exploration of the FEL model in related circumstances, such as ion-cyclotron waves in the magnetosphere, or toroidal configurations, where the excitation of whistler waves by runaway electrons has recently been observed (Spong et al., 2018).

## 4. Conclusion

In this work, we investigated nonlinear aspects of whistler-mode chorus amplification in the magnetosphere using the FEL model. In the single-mode case, we derived nonlinear collective variable equations for the system, predicting exponential growth followed by saturation and post-saturation amplitude oscillations consistent with observational data. Next, we considered a packet of whistlers with a spectrum of frequencies and spatially dependent amplitudes, and found that the wave behavior is approximately governed by the complex GLE which admits solitary wave solutions. We found that both the single-mode and multi-mode equations predict amplitude modulations on millisecond timescales, consistent with observations. Further exploration into the FEL model, including the newly proposed GLE, its stability, and its observational implications, will be the subject of a future publication.

## Conflict of Interest

The authors declare no conflicts of interest relevant to this study.

## Data Availability Statement

Data were not used, nor created for this research. The figures were produced with Wolfram Mathematica version 14.2.1.0, using a small number of standard functions to numerically solve the equations presented in the manuscript.

**Acknowledgments**

B. Bonham gratefully acknowledges the unwavering mentorship and support of Professor Lyman Page, countless discussions with Erin Phillips whose perspectives have greatly enriched his understanding, and insightful mathematical discussions with Shouda Wang. We also thank the anonymous reviewers for their thoughtful suggestions which helped improve our presentation. Both authors acknowledge support from the Simulation Center of electrons (SCREAM) SciDAC center by Office of Fusion Energy Science and Office of Advanced Scientific Computing of U. S. Department of Energy, under contract No. DE-SC0016268 and DE-AC02-09CH11466 as well as NSF Award No. 2209471. B. Bonham also acknowledges generous support from the Princeton University Department of Physics made possible by Department Chair Jim Olsen.

## Erratum

The originally published version of this article contained an error in the equation that appears at the end of the second sentence of the fifth paragraph of Section 2.1. The equation has been corrected as follows: $\Gamma_n \equiv \left.\frac{d^n\gamma}{d\eta_z^n}\right|_{\eta_{z0}}$ This may be considered the authoritative version of record.